\input{aipcheck}

\documentclass[
    ,final            
  ]
  {aipproc}

\layoutstyle{8x11double}

\begin{document}

\title{A radio and near-infrared mini-survey of the MGRO~J2019+37 complex}


\author{Paredes, J.M.}{
  address={
Dept. d'Astronomia i Meteorologia and Institut de Ci\`encies del Cosmos (ICC),
Universitat de Barcelona, Mart\'{\i} i Franqu\`es, 1, 08028 Barcelona, Spain
}
}

\author{Mart\'{\i}, J.}{
  address={
Grupo de Investigaci\'on FQM-322, 
Universidad de Ja\'en, Campus Las Lagunillas s/n, A3, 23071 Ja\'en, Spain
}
}

\author{Ishwara-Chandra, C.H.}{
  address={National Centre for Radio Astrophysics, TIFR,
Post Bag 3, Ganeshkhind, Pune-411 007, India 
}
}

\author{S\'anchez-Sutil, J.R.}{
  address={
Grupo de Investigaci\'on FQM-322, 
Universidad de Ja\'en, Campus Las Lagunillas s/n, A3, 23071 Ja\'en, Spain
}
}

\author{Mu\~noz-Arjonilla, A. J.}{
  address={
Grupo de Investigaci\'on FQM-322, 
Universidad de Ja\'en, Campus Las Lagunillas s/n, A3, 23071 Ja\'en, Spain
}
}

\author{Zabalza, V.}{
  address={
Dept. d'Astronomia i Meteorologia and Institut de Ci\`encies del Cosmos (ICC),
Universitat de Barcelona, Mart\'{\i} i Franqu\`es, 1, 08028 Barcelona, Spain
}
}

\author{Rib\'o, M.}{
  address={
Dept. d'Astronomia i Meteorologia and Institut de Ci\`encies del Cosmos (ICC),
Universitat de Barcelona, Mart\'{\i} i Franqu\`es, 1, 08028 Barcelona, Spain
}
}

\author{Luque-Escamilla, P. L.}{
  address={
Grupo de Investigaci\'on FQM-322, 
Universidad de Ja\'en, Campus Las Lagunillas s/n, A3, 23071 Ja\'en, Spain
}
}

\author{Bosch-Ramon, V.}{
  address={
 Max Planck Institut f\"ur Kernphysik, Saupfercheckweg 1, Heidelberg 69117, Germany
}
}

\author{Peracaula, M.}{
  address={Institut d'Inform\'atica i Aplicacions, Universitat de Girona, Girona, Spain}
}

\author{Mold\'on, J.}{
  address={
Dept. d'Astronomia i Meteorologia and Institut de Ci\`encies del Cosmos (ICC),
Universitat de Barcelona, Mart\'{\i} i Franqu\`es, 1, 08028 Barcelona, Spain
}
}

\author{Bordas, P.}{
  address={
Dept. d'Astronomia i Meteorologia and Institut de Ci\`encies del Cosmos (ICC),
Universitat de Barcelona, Mart\'{\i} i Franqu\`es, 1, 08028 Barcelona, Spain
}
}

\begin{abstract}

MGRO~J2019+37 is an unidentified source of very high energy gamma-rays
originally reported by the MILAGRO collaboration as the brightest TeV source in
the Cygnus region. Despite the poor angular resolution of MILAGRO, this object
seems to be most likely an extended source or, alternatively, a superposition of
point-like TeV sources.

In order to contribute to the understanding of this peculiar object, we have
mosaiced it with the Giant Metrewave Radio Telescope (GMRT) in Pune, India, at
the 610~MHz frequency covering a field of view of about 6 square degrees down
to a typical rms noise of a few tenths of mJy. We also observed the central
square degree of this mosaic in the near infrared ${\it K_{\rm s}}$-band using the 3.5~m
telescope and the OMEGA2000 camera at the Calar Alto observatory (Spain).  We
present here a first account of our observations and results, together with a
preliminar cross correlation of the radio and infrared source catalog as well as
a correlation with the available X-ray observations of the region.

\end{abstract}

\maketitle


\section{MGRO~J2019+37}

The galactic very-high-energy (VHE) $\gamma$-ray sources discovered during these
last years by the new generation of \^Cerenkov telescopes (H.E.S.S., MAGIC, MILAGRO)
are currently a hot topic in modern high-energy Astrophysics. Among the $\sim$75
detected sources, nearly one third of them still remain unidentified.

A recent addition to the population of extended and unidentified TeV sources
has been reported by the MILAGRO collaboration, with the discovery in the Cygnus
region of the most extended TeV source ever known \citep{abdo07}. The TeV
emission from this area covers several square degrees and includes diffuse
emission and at least a new source, MGRO J2019+37. The origin of these emissions
and their association to any astrophysical counterpart is unknown.  Although a
possible connection with anisotropy of the galactic cosmic rays has been  proposed \citep{amenomori}, the TeV $\gamma$-ray flux as measured at
12~TeV from the diffuse emission of the Cygnus region (after excluding MGRO
J2019+37) exceeds that predicted from a conventional model of cosmic ray
production and propagation \citep{abdo07}.  This strongly suggests the existence
of hard-spectrum cosmic-ray sources and/or other types of TeV $\gamma$-ray
sources in the region.

To better understand this peculiar object, we have performed a multi-wavelength
campaign comprising a radio survey at 610 MHz and infrared observations in the
${\it K_{\rm s}}$ band. Recent X-ray observations are being analyzed.

\begin{figure*}
      \includegraphics[width=\textwidth]{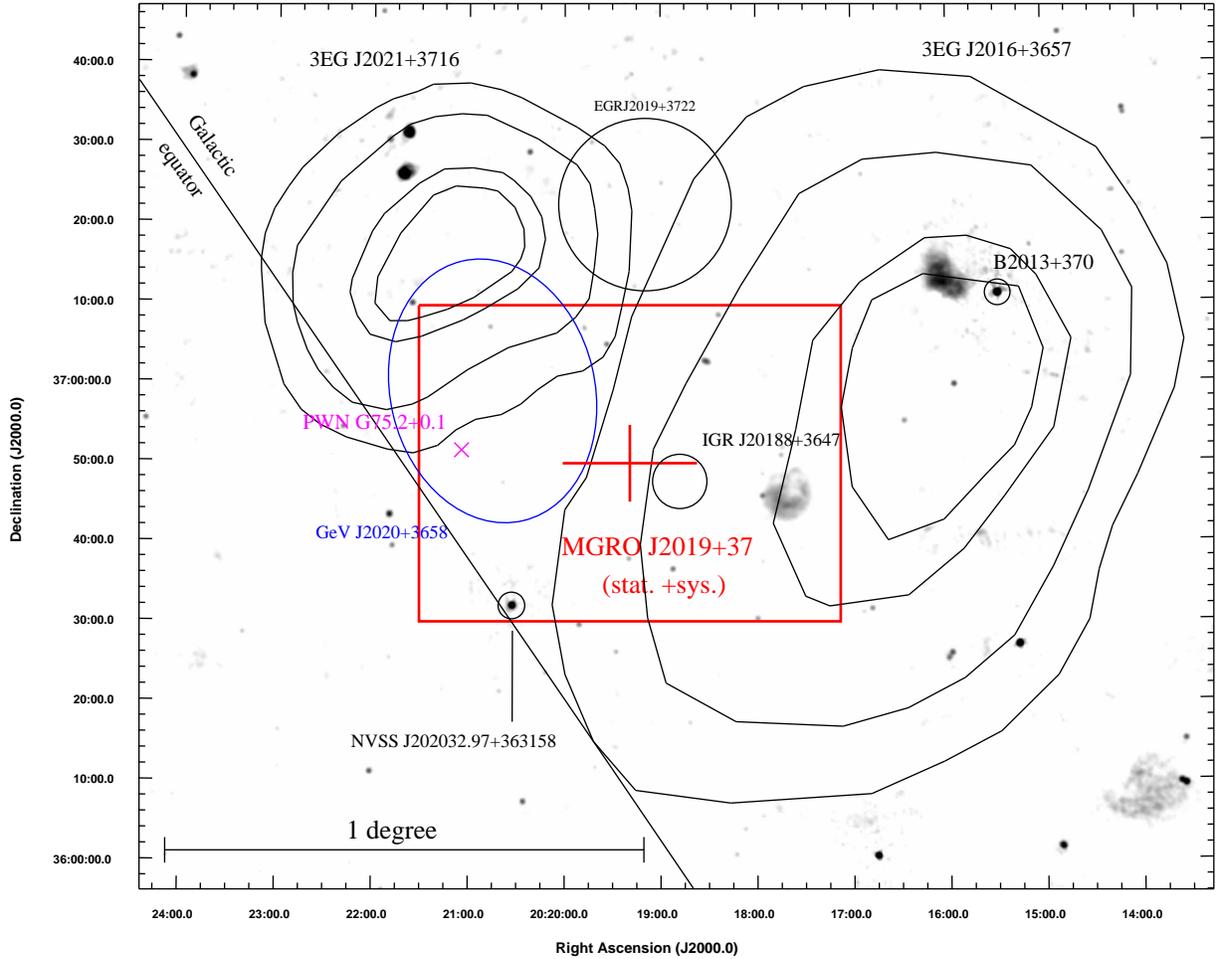}
      \caption{Radio map obtained with the GMRT radiotelescope at 610 MHz
      (greyscale) convolved with a circular restoring beam of 30~arcsecond.
      The red cross and box indicate the Center of Gravity and
      uncertainty region of the TeV emission from the source MGRO~J2019+37.
The conspicuous radio sources located inside this box correspond to an extended HII region
and a bright compact radio source also detected with the VLA as NVSS J202032.97+363158.
      Overlayed are the position probability contours (50\%, 68\%, 95\% and
      99\%) of the Third EGRET Catalog sources 3EG~J2021+3716 and 3EG~J2016+3657
      \citep{hartman}, as well as the GeV source GeV J2020+3658 (blue ellipse)
      \citep{gev}. Recently EGRET data has been reanalyzed by \cite{egr},
      resulting in a single source named EGR J2019+372, located approximately between the two 3EG sources.}
\end{figure*}

\section{Radio Survey}

The region has been observed with wide-field deep radio imaging at 610~MHz
(50~cm) using the GMRT in Pune (India) on July 2007. A total of 19 pointings cover the region
of about $2.5^\circ\times2.5^\circ$ centered on the MGRO~J2019+37 peak of
emission. The final combined image has an rms of
$\sim$0.2~mJy/beam with a 5~arcsecond resolution thanks to the long baselines of the
GMRT antennas. Standard flagging and calibration procedures were used to process the GMRT data
using the AIPS software package.
For illustration purposes, in Fig. 1 we show a low-resolution version
of the GMRT mosaic produced using a restoring beam of 30~arcsecond in order to better
enhance the extended radio sources in the field.

The AIPS task SAD was applied for source extraction in the high-angular resolution version of Fig. 1.
As a result, a total of 
358 radio sources were detected above 2~mJy. 
Among them, 255 are fainter than 10~mJy and the majority were previously undetected at radio wavelengths.
In Fig.~2 we show their distribution histogram as a
function of $\mbox{log}\ S_\nu^{\rm Peak}$.  Apart from the already known
extended bright sources, we have been able to detect new extended mJy sources
displaying arcsecond structure.  Further studies on the relevant radio sources
are being conducted to explore possible counterparts or particle accelerators
related to the TeV source.



\section{Infrared Survey}

We have also carried out an infrared survey of the central square degree of the
region using the OMEGA2000 wide field camera on the 3.5~m telescope at the
Centro Astron\'omico Hispano Alem\'an (CAHA) in Almer\'{\i}a (Spain) on 25 September 2007. 
This instrument consists of a Rockwell HAWAII2 HgCdTe
detector with $2048\times2048$ pixels sensitive within 
0.8 to 2.5 $\mu$m.
A pattern of $4 \times 4$ pointings was performed in the ${\it K_{\rm s}}$-band (2.15 $\mu$m) in order to
minimize the interstellar absorption. Individual frames were
sky-subtracted, flat-field corrected and then combined into a final mosaic
using the AIPS task FLATN. A low-resolution view of this mosaic is presented in Fig. 3.
A catalog of $\sim 3\times10^{5}$ near infrared sources was produced using
the SEXtractor package.



\section{Catalog Cross-Correlation}

We have performed a preliminar cross-correlation of the radio and infrared
source catalogs. Out of the 358 detected radio sources, 40 of them lie inside
the central squared degree imaged in the near infrared. A total of 7 of these sources have an
infrared counterpart within less of 1 arcsecond of their radio position.

We have also obtained source lists of all X-ray observations of the region
performed by {\it Chandra} and {\it XMM-Newton}, computed through the \verb+celldetect+
and \verb+edetect_chain+ tasks. A total of 22 of the 358 radio sources are located in
fields observed in X-rays, which cover an area of $\sim 4\times10^3
\mbox{arcmin}^2$. We have found that 5 of them have a radio counterpart within 5
arcseconds. Extrapolating the fraction of X-ray counterparts to the rest of the
radio observation, we expect that about 82 of the detected radio sources have an
X-ray counterpart.

\begin{figure*}
      \includegraphics[width=\columnwidth]{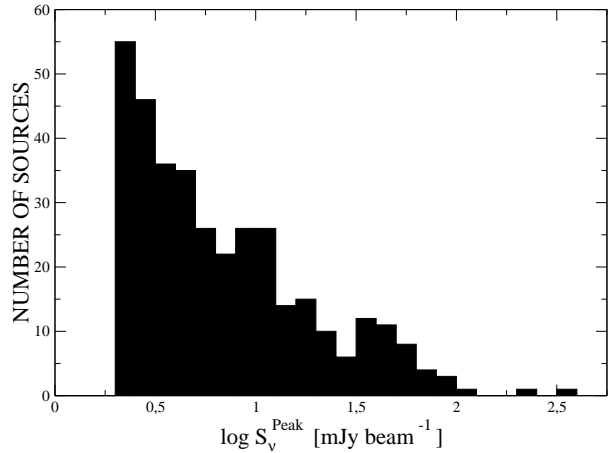}
      \caption{Number of sources vs.~$\mbox{log}\ S_\nu^{\rm Peak}$ for the 358
      sources detected in the GMRT 610~MHz radio mosaic.}
\end{figure*}

\begin{figure*}
      \includegraphics[width=\textwidth]{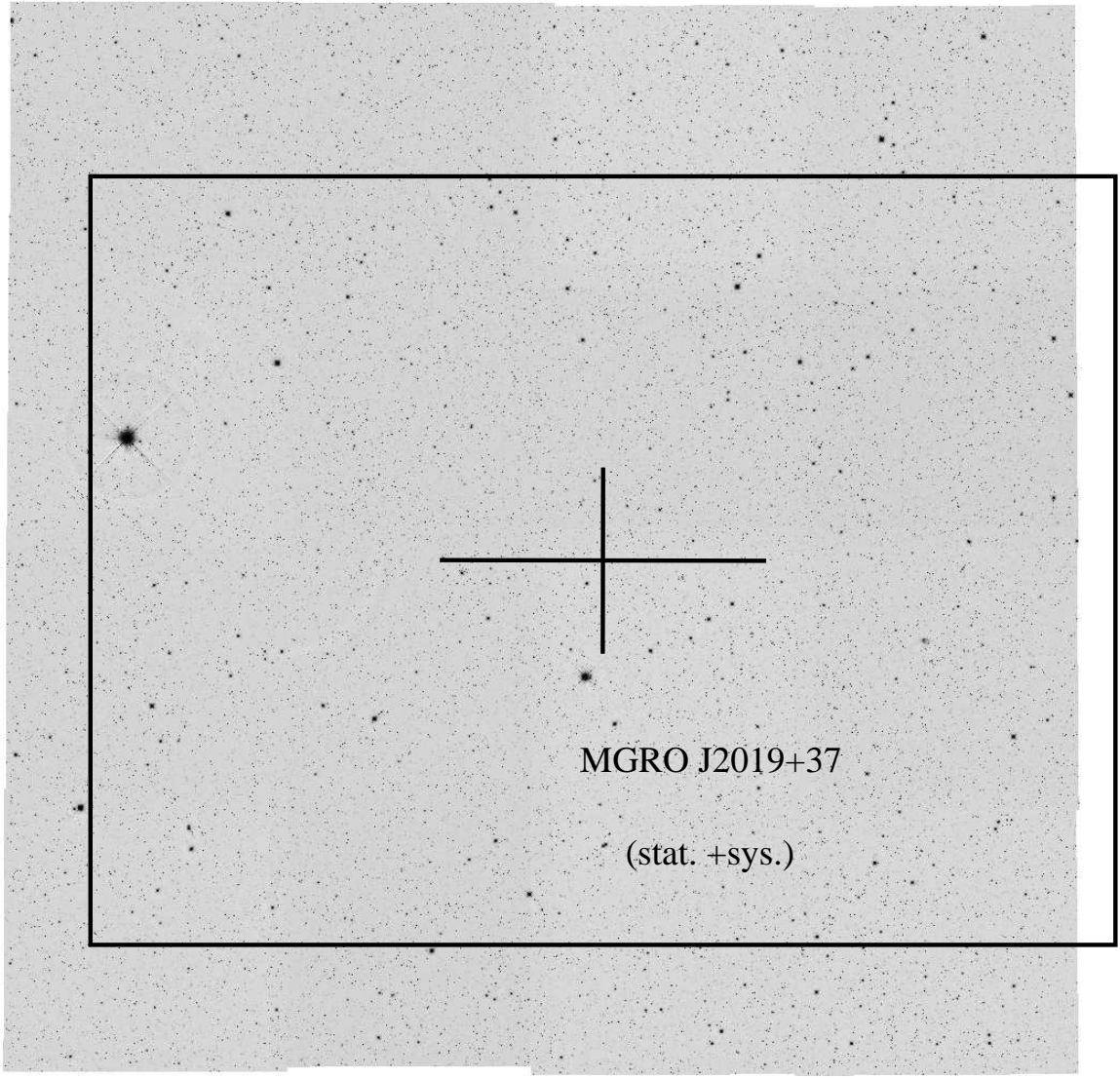}
      \caption{Appearance of the full near infrared mosaic in the ${\it K_{\rm s}}$-band
obtained with the CAHA 3.5~m telescope and the OMEGA2000 camera. The ensamble
of 16 pointing covers almost completely the Center of Gravity and
uncertainty region of the TeV emission from the source MGRO~J2019+37.
The limiting magnitude is ${\it K_{\rm s}}\sim 17$ and the total field of view is
about 1 squared degree. North is up and East is to the left.
Astrometric solutions on the final frames were determined within $\pm$0.1 arc-second by identifying
about twenty reference stars in each pointing for which positions were retrieved
from the 2MASS catalog.
}
\end{figure*}

\section{Concluding Remarks}

The analysis of both the infrared and radio observations has very recently been
finished. Having this kind of data for the 
MGRO~J2019+37 field will be a very useful information by the time the {\it Fermi}
satellite provides more accurate information about the gamma-ray sources in
this remarkable region of the Galactic Plane.
The physical understanding of the most relevant sources in the field
is currently work in progress, as well as the analysis of new {\it XMM-Newton} and
{\ AGILE} observations.


\begin{theacknowledgments}

We acknowledge support by DGI of the Spanish Ministerio de
Educaci\'on y Ci\-encia (MEC) under grants AYA2007-68034-C03-01,
AYA2007-68034-C03-02 and AYA2007-68034-C03-03, FEDER funds and Junta de
Andaluc\'{\i}a under PAIDI research group FQM-322. GMRT is run by the National
Centre for Radio Astrophysics of the Tata Institute of Fundamental
Research. This publication makes
use of data products from the Two Micron All Sky Survey, which is a
joint project of the University of Massachusetts and the Infrared
Processing and Analysis Center/California Institute of Technology,
funded by the NASA and NSF in the USA.


\end{theacknowledgments}





\IfFileExists{\jobname.bbl}{}
 {\typeout{}
  \typeout{******************************************}
 \typeout{** Please run "bibtex \jobname" to optain}
  \typeout{** the bibliography and then re-run LaTeX}
  \typeout{** twice to fix the references!}
  \typeout{******************************************}
  \typeout{}
 }



\end{document}